\documentclass[transmag]{IEEEtran}

\usepackage{styles/spconf}
\usepackage{latexsym}
\usepackage{graphicx}
\usepackage{amsmath,amssymb,latexsym,float,epsfig,subfigure,overpic}
\usepackage{paralist}
\usepackage{multirow}
\usepackage{siunitx}
\usepackage{adjustbox}
\usepackage{amsmath}
\usepackage{hyperref}
\usepackage{color}

\DeclareMathOperator*{\argmin}{arg\,min}

\def\BibTeX{{\rm B\kern-.05em{\sc i\kern-.025em b}\kern-.08em T\kern-.1667em\lower.7ex\hbox{E}\kern-.125emX}}
\title{Adaptive Compressive Sampling for Mid-infrared Spectroscopic Imaging}
\name{M. Lotfollahi,$^{1a}$ N. Tran,$^1$ C. Gajjela,$^1$ S. Berisha,$^2$ Z. Han,$^1$ D. Mayerich,$^1$ R. Reddy.$^1$\thanks{This work was supported in part by the NLM Training Program in Biomedical Informatics and Data Science T15LM007093 (RM, RR), the Cancer Prevention and Research Institute of Texas (CPRIT) \#RR170075 (RR), National Institutes of Health \#R01HL146745 (DM), and the National Science Foundation CAREER Award \#1943455 (DM).} \thanks{$^a$ Current Address: $3$M Corporate Research Analytical Laboratory, St. Paul, MN} }

\address{$^1$University of Houston, Houston TX, $^2$Milwaukee School of Engineering, Milwaukee WI}

\begin{document}
\maketitle
\begin{abstract}
Mid-infrared spectroscopic imaging (MIRSI) is an emerging class of label-free, biochemically quantitative technologies targeting digital histopathology. Conventional histopathology relies on chemical stains that alter tissue color. This approach is qualitative, often making histopathologic examination subjective and difficult to quantify. MIRSI addresses these challenges through quantitative and repeatable imaging that leverages native molecular contrast. Fourier transform infrared (FTIR) imaging, the best-known MIRSI technology, has two challenges that have hindered its widespread adoption: data collection speed and spatial resolution. Recent technological breakthroughs, such as photothermal MIRSI, provide an order of magnitude improvement in spatial resolution. However, this comes at the cost of acquisition speed, which is impractical for clinical tissue samples. This paper introduces an adaptive compressive sampling technique to reduce hyperspectral data acquisition time by an order of magnitude by leveraging spectral and spatial sparsity. This method identifies the most informative spatial and spectral features, integrates a fast tensor completion algorithm to reconstruct megapixel-scale images, and demonstrates speed advantages over FTIR imaging while providing spatial resolutions comparable to new photothermal approaches.
\end{abstract}


\section{INTRODUCTION}
Mid-infrared spectroscopic imaging (MIRSI) is an emerging class of quantitative label-free molecular identification technologies impacting several fields, including histopathology  \cite{pahlow2018application}, material science \cite{xu2019ftir}, environmental and toxicological chemistry \cite{trevisan2012extracting}, and forensics \cite{sharma2019ftir}. Fourier transform infrared (FTIR) spectroscopic imaging is the best-known MIRSI technique that provides spatially resolved images of molecular constituents. FTIR imaging probes the sample with mid-infrared light (\numrange{2.5}{12.5}\si{\micro\meter}), and has a wavelength $\lambda$ dependant, diffraction-limited  spatial resolution \cite{reddy2013high}. While this resolution is sufficient for numerous applications, it limits the applicability of MIRSI in several fields, such as cellular and sub-cellular microscopy, where features are smaller than $\approx$\SI{6}{\micro\meter} in size. 

New MIRSI technologies such as photothermal infrared (PTIR) \cite{bialkowski1996photothermal} and optical photothermal infrared (O-PTIR) \cite{zhang2016depth} spectroscopy overcome the FTIR diffraction limit using a pump-probe instrumentation architecture. The sample is excited using a mid-infrared quantum cascade laser source, and localized absorbance is measured using an atomic force microscope \cite{bialkowski1996photothermal} or secondary optical beam with a smaller resolution \cite{zhang2016depth}. These techniques provide sub-micrometer resolution but are slow at obtaining hyperspectral data on large samples. For example, O-PTIR imaging of a \SI{1}{\milli\meter^2} sample at \SI{250}{\nano\meter} pixel spacing, which is typical in histology, requires approximately 46 days using current instrumentation.

We hypothesize that acquisition time can be dramatically reduced using compressive sampling \cite{candes2008introduction} to take advantage of spatial and spectral sparsity inherent in MIRSI data \cite{deutsch2015compositional}. Instrumentation for O-PTIR is versatile and allows for the acquisition of (1) a 1D spectrum or (2) a 2D band image at one wavenumber in one measurement. We will utilize this versatility along with sparsity-based reconstructions to achieve an order of magnitude improvement in data collection speed.   

This paper describes an adaptive method to identify the most informative spatial and spectral features given the O-PTIR sampling constraints. We then simulate data acquisition on large FTIR images and reconstruct the resulting hyperspectral image (HSI) using a fast tensor completion algorithm to handle megapixel-scale hyperspectral data. This allows us to directly compare our reconstruction results to the underlying ground truth data and validate our algorithm.  We also demonstrate that the most common metrics used to quantify reconstructed HSI quality are inadequate \textit{in practice}. We, therefore, propose a classification-based metric compatible with FTIR image assessment.
\section{Compressive Sensing}
Compressive sensing (CS) is extensively employed in digital image and signal processing \cite{ender2013brief,willett2011compressed,chan2008terahertz}  to reduce data throughput by leveraging sparse optimization to reconstruct incomplete measured data \cite{candes2008introduction}. CS theory  \cite{candes2005l1} utilizes the fact that objects being imaged have a sparse representation in some
basis. FTIR spectra are sparse when projected on a principal component \cite{reddy2010accurate} or Fourier bases \cite{nguyen2012denoising}. 

In this paper, we reconstruct a large hyperspectral data cube with the size of $N_S\times Z_S$ using $N_D$ selected point spectra ($N_D << N_S$) and $Z_D$ spectral band images ($Z_D<<Z_S$) selected in different wavenumbers, where $N_S = X_S \times Y_S$ is the number of pixels in the spatial domain and $Z_S$ is the number of spectral bands in the reconstructed hyperspectral data. 
\section{Adaptive Sampling}
Our approach uses an iterative method of adaptive sampling and image reconstruction. We identify and collect the most informative band images and point spectra at each iteration. This new data is then integrated into a reconstruction used to inform sampling in the next iteration.

\begin{figure*}
    \centering
    \includegraphics[width=\linewidth]{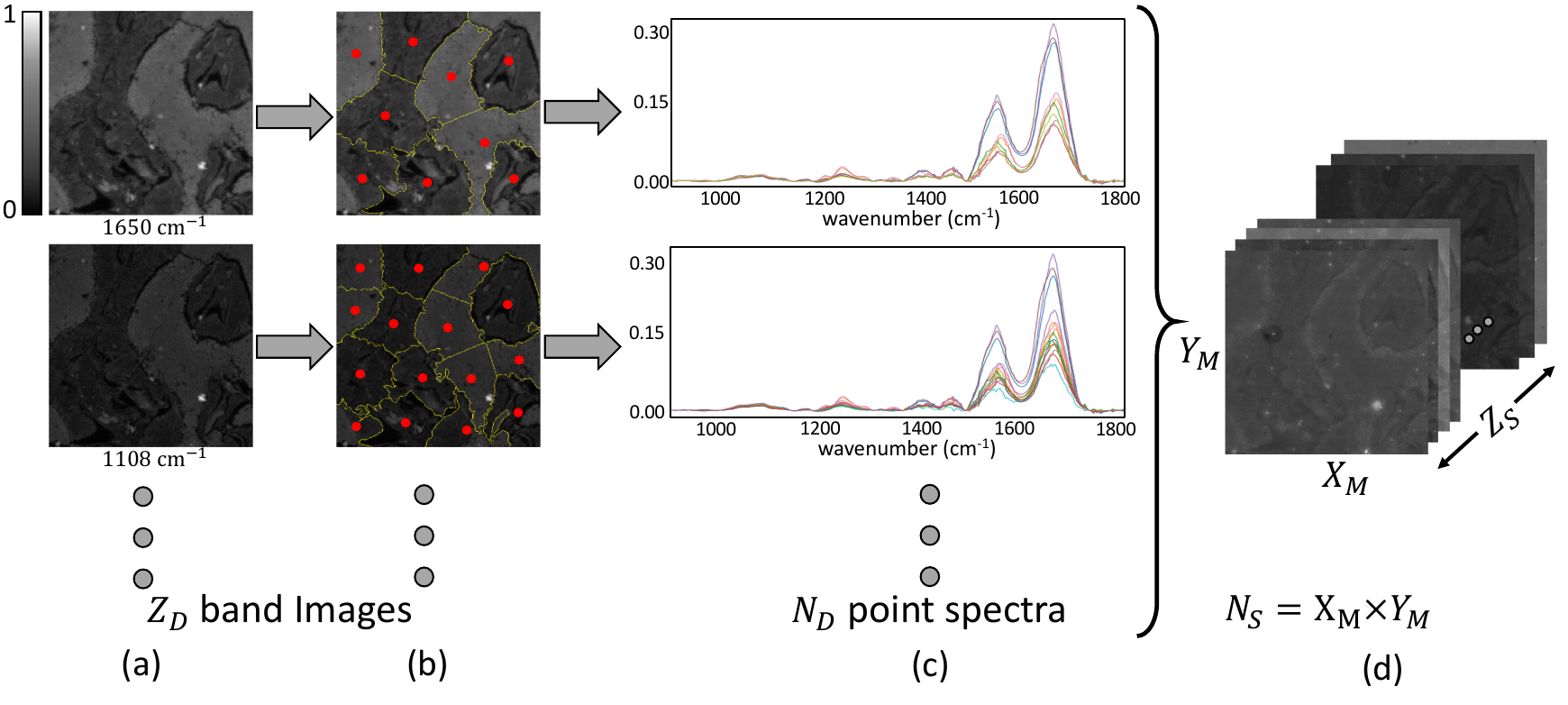}
    \caption{Schematic representation of adaptive sampling using the iterative method. (a) Selected band images. (b) Clustered images using SLIC, the center of superpixels are specified as red circles. (c) IR spectra at the center of superpixels. (d) By integrating $Z_D<<Z_S$ band images and $N_D<<N_S$ point spectra, a high-resolution hyperspectral image ($N_S \times Z_S$) is reconstructed using the proposed method.}
    \label{fig:ad_sampling}
\end{figure*}

We present an iterative method to select point spectra and band images for optimally reconstructing the hyperspectral volume. We initiate the process by taking the band image at Amide I (\SI{1654}{\per\centi\meter}), which provides the most molecular information for biological samples (Fig.\ \ref{fig:ad_sampling}a, first row). We then apply Simple Linear Iterative Clustering (SLIC) \cite{achanta2010slic} to perform a local clustering of pixels. SLIC clusters pixels into the desired number of superpixels ($N$) based on their color, similarity, and proximity in the image plane (Fig.\ \ref{fig:ad_sampling}b, first row). Point spectra are measured at the centers of superpixels (Fig.\ \ref{fig:ad_sampling}c, first row).
We select a new band image and new point spectra in an alternating manner (Fig.\ \ref{fig:ad_sampling}a-c, second row). After multiple iteration, a high-resolution hyperpectral image is reconstructed using the method described in section \ref{sec:recon}(Fig.\ \ref{fig:ad_sampling}d).

The band image for the second iteration is chosen based on the first band image and the set of measured point spectra. We compute the spectral correlation coefficient between the first band and all bands from the measured point spectra. The band with the lowest correlation coefficient is the most informative band. Noise is also uncorrelated with the first band image, but has a low magnitude. Therefore, the optimal band is chosen to have the lowest correlation with the first band and the highest magnitude. Having obtained a second band image, the SLIC algorithm is applied to find the next set of points to measure spectra. The iterative band selection process is initialized with a pair of bands $B_1= 1654 \ \mathrm{ cm^{-1}}$ and $B_2$, followed by the orthogonal subspace projection (OSP) algorithm \cite{du2008similarity}. Before applying OSP, the bands with low SNR and the ones corresponding to water absorption should be removed. Those bands can be selected by computing the spectral correlation coefficient matrix between the original bands. The bands with low correlation with adjacent bands (low value in upper diagonal of correlation coefficient matrix) carry unrelated information (noise) and should be excluded. 

The initial set of selected sub-bands is defined as $\Phi_0 = \{B_1,B_2\}$. The next band $B_3$ should provide the maximum amount of additional information and is therefore selected to be the most \textit{dissimilar} to $\Phi_0$. $B_3$, therefore, has the maximum projection on the orthogonal subspace of $B_1$ and $B_2$.
$$
\mathbf{P} = \mathbf{I} - \mathbf{Z}(\mathbf{Z}^\intercal \mathbf{Z})^{-1} \mathbf{Z}^\intercal,
$$
where $\mathbf{Z}$ is an $N\times 2$ matrix where the $N$-pixel columns represent the $\sqrt{N}\times\sqrt{N}$ band images $B_1$ and $B_2$ respectively. The projection of all selected points ($y ^{N\times 1}$) in band ($B$) on the orthogonal subspace of $\Phi$ are computed as
$$
y_0 = \mathbf{P}^\intercal y.
$$
The band with the maximum orthogonal component $\left\|y_0\right\|$ is the most dissimilar, therefore $\Phi$ is updated to include the new band image: $\Phi_1=\{B_1,B_2,B_3\}$. This process continues after selecting additional point spectra to exploit all available information. New spectral bands are selected iteratively until the desired threshold is reached.

\section{Reconstruction} \label{sec:recon}
We use a hyperspectral fusion algorithm with a LASSO optimizer \cite{2020arXiv200306944T} for reconstruction. This algorithm requires two inputs to generate a high-resolution hyperspectral image. The first input $\mathbf{H} \in \mathbb{R}^{ N_D \times Z_S}$ consists of full spectra at $N_D$ selected points. The second input $\mathbf{M} \in \mathbb{R}^{ N_S \times Z_D}$ consists of images at selected spectral bands. Our goal is to leverage a small number of spectral and spatial samples, such that $N_D \ll N_S$ and $Z_D \ll Z_S$, to generate a fusion image $\mathbf{S} \in \mathbb{R}^{N_S \times Z_S}$ with high resolution in \textit{both} spectral and spatial domains.
\subsection{Image Generation}
Image $\mathbf{H}$ is represented as:
\begin{equation}
    \mathbf{H} = \mathbf{L}\mathbf{S} + \boldsymbol{\eta}_1,
    \label{eqn:HI}
\end{equation}
where $\mathbf{L} \in \mathbb{R}^{N_D \times N_S}$ is an operator to re-sample $\mathbf{S}$. Similarly, image $\mathbf{M}$ is generated by:
\begin{equation}
    \mathbf{M} = \mathbf{S}\mathbf{B} + \boldsymbol{\eta}_2,
    \label{eqn:MI}
\end{equation}
where $\mathbf{B} \in \mathbb{R}^{Z_S \times Z_D}$ is a sparse matrix that extract individual bands from $\mathbf{S}$. Both $\eta_1$ and $\eta_2$ are assumed to be additive white Gaussian noise.
The next step is PCA where the fusion image is defined as $\mathbf{S}= [s_1,...,s_{N_S}]^T$. Each rows of vector $\mathbf{s}_i = [\mathbf{s}_{i,1},\mathbf{s}_{i,2}, ...,\mathbf{s}_{i,Z_S}]$ contains all the spectral information at a spatial pixel location. We represent $\mathbf{s}_i$ in subspace $\mathbf{Q}$ as:
\begin{equation}
    \mathbf{s_i} = \mathbf{r}_i\mathbf{Q},
    \label{eqn:FI}
\end{equation}
where $\mathbf{r}_i \in \mathbb{R}^{1 \times \tilde{Z}}$ is the projection of $\mathbf{s}_i$ onto the orthonormal subspace spanned by the columns of $\mathbf{Q} \in \mathbb{R}^{\tilde{Z} \times Z_S}$, such that $\tilde{Z} \ll Z_S$. The image $\mathbf{R} \in \mathbb{R}^{N_S \times \tilde{Z}}$ projected onto this subspace is composed of the reduced spectra: $\mathbf{R} = [\mathbf{r}_1, ..., \mathbf{r}_{N_S}]^T$. Integrating (Eqn.\ \ref{eqn:FI}) into (Eqn.\ \ref{eqn:HI}) and (Eqn.\ \ref{eqn:MI}), we can express $\mathbf{H}$ and $\mathbf{M}$ as:
\begin{equation}
    \mathbf{H} = \mathbf{L}\mathbf{R}\mathbf{Q} + \mathbf{\eta}_1,
    \label{eqn:subspaceHS}
\end{equation}
\begin{equation}
    \mathbf{M} = \mathbf{R}\mathbf{Q}\mathbf{B} + \mathbf{\eta}_2.
    \label{eqn:subspaceMS}
\end{equation}
\subsection{Initialization}
$\mathbf{\bar{R}}$ is the initialization of $\mathbf{R}$ using Maximum a Posteriori (MAP) and it is defined as:
\begin{multline}
    \mathbf{\bar{R}} =
    \Big(\mathbf{Q}^T\mathbf{L}^T\mathbf{\Lambda_H}\mathbf{L}\mathbf{Q}+ \mathbf{\Lambda_{\bar{R}|M}}\Big)^{-1}\\
    \Big(\mathbf{L}^T\mathbf{\Lambda_H}\mathbf{Q}^T\mathbf{H} + \mathbf{\Lambda_{\bar{R}|M}}^{-1}\mathbf{\tilde{R}}\Big).
    \label{eqn:RbarMAP}
\end{multline}
where
\begin{itemize}
    \item $\mathbf{\tilde{R}}$ $= \mathbb{E}\{\mathbf{\bar{R}}|\mathbf{M}\}$ is the expected value of $\mathbf{\bar{R}}$ given $\mathbf{M}$,
        \item Rewrite as: {$\mathbf{\tilde{R}}$} $= \mathbb{E}(\mathbf{\bar{R}}) + \frac{\mathbf{\Lambda}_{\mathbf{\bar{R}},\mathbf{M}} [\mathbf{M} - \mathbb{E}(\mathbf{M})]}{\mathbf{\Lambda}_{\mathbf{M,M}}}$,
            \item $\mathbf{\Lambda}_{\mathbf{\bar{R}},\mathbf{M}}$ is the cross-covariance matrices with form:\\ $\mathbf{\Lambda}_{\mathbf{\bar{R}},\mathbf{M}} =\mathbb{E}\big[ \big(\mathbf{\bar{R}} - \mathbb{E}(\mathbf{M})\big)\big(\mathbf{\bar{R}} - \mathbb{E}(\mathbf{M})\big)^T\big],$
    \item $\mathbf{\Lambda}_{\mathbf{\bar{R}}|\mathbf{M}}$ is the row covariance matrix of $\mathbf{R}$ given $\mathbf{M}$ as:
    $\mathbf{\Lambda}_{\mathbf{\bar{R}}|\mathbf{M}} = \mathbf{\Lambda}_{\mathbf{\bar{R}},\mathbf{\bar{R}}} - \frac{\mathbf{\Lambda}_{\mathbf{\bar{R}},\mathbf{M}}\mathbf{\Lambda}^T_{\mathbf{\bar{R}},\mathbf{M}}}{\mathbf{\Lambda}_{\mathbf{M},\mathbf{M}}}$.
\end{itemize}
\subsection{Least Absolute Shrinkage and Selection Operator (LASSO)}
In the next step, LASSO is implemented to solve for $\mathbf{R}$ in Equation (\ref{eqn:subspaceHS}) and (\ref{eqn:subspaceMS}). The optimization equation is as follow:
\begin{multline}
    \argmin_{\mathbf{R}} \frac{1}{2} \left\lVert\mathbf{\Lambda_H}^{-\frac{1}{2}}(\mathbf{H} - \mathbf{LRQ})\right\rVert_F^2\\
    + \frac{1}{2}\left\lVert\mathbf{\Lambda_M}^{-\frac{1}{2}}(\mathbf{M} - \mathbf{RQB})\right\rVert_F^2
    + \eta\left\lVert\mathbf{R}\right\rVert_n.
    \label{eqn:MaxaP}
\end{multline}
In the final step, ADMM is implemented to solve Equation \ref{eqn:MaxaP} \cite{2020arXiv200306944T}.

\section{Results}
The proposed reconstruction method is applied to FTIR hyperspectral data in high magnification mode from TMA BR961 using the Agilent Stingray imaging system. The imaging system is composed of a Cary 680 spectrometer coupled to a Cary 620 microscope with $0.62$ numerical aperture, projected pixel size of \SI{1.1}{\micro\meter} and spectral resolution of \SI{8}{\centi\meter ^{-1}}. We truncated the spectral range to the fingerprint wavenumbers between \numrange{900}{1800}\si{\per\centi\meter}, the same as the one used in photothermal spectroscopy. 

To reconstruct a single core with the size of $1380\times1380\times117$ from the TMA96, our algorithm only requires a few high-resolution band images and a limited set of point spectra. The process begins with acquiring the first band image at \SI{1654}{\per\centi\meter}. In each iteration, a new band image is added based on the band selection process described in the "Adaptive Sampling" section. Then, SLIC is applied to the available band images to extract the desired number of superpixels. The centers of superpixels are the positions of point spectra. We initialize the number of point spectra in each iteration to $20,40,80,160,240,480$. However, the number of superpixels generated is adaptive and deviates from the initialized numbers since we update the initialized centers to the lowest gradient position in corresponding local neighborhoods \cite{wang2017bslic}. A Gaussian smoothing kernel with $\sigma=5$ is applied prior to clustering, and the compactness factor is set to $0.03$. 

We compare the original and reconstructed image in each iteration using multiple performance metrics, including peak signal-to-noise ratio (PSNR), Root Mean Square Error (RMSE), Spectral Angle Mapper (SAM), and Relative Dimensionless Global Error (ERGAS)  \cite{wald2000quality,jagalingam2015review,sara2019image,kavitha2016survey}. 
In addition, to ensure that the reconstructed data can capture all biological features, support vector machines (SVM) with radial basis function (RBF) kernel is used. RBF SVM has been validated extensively on single-pixel MIRSI spectra \cite{berisha2019deep} and used here to identify six major cellular and acellular constituents of tissue: adipocytes, blood, collagen, epithelium, necrosis, and myofibroblasts. The RBF SVM classifier is trained using $10,000$ pixels per class from TMA BR961 \cite{berisha2019deep}. RBF SVM provides overall accuracy of $94\%$ on a test FTIR breast core. Experimentally measured raw HSI data from this core is also used to validate the reconstruction.

\begin{table}[tb]

\caption{Performance metrics and classification accuracy in each iteration by adding one band image and given number of point spectra.}
\resizebox{\columnwidth}{!}{\begin{tabular}{|c|c|l|c|c|c|c|c|}
\hline

\multirow{2}{*}{\#Iter} & \multirow{2}{*}{\#Point Spectra} & \multicolumn{1}{c|}{\multirow{2}{*}{Selected Wavenumbers}} & \multicolumn{4}{c|}{Performance Metrics} & RBF SVM \\ \cline{4-7}

                        &                                  & \multicolumn{1}{c|}{}                                      & PSNR     & RMSE     & SAM      & ERGAS    & Accuracy       \\ \hline
1                       & 14                               & 1654                                                       & 21.86    & 0.0806   & 2.1465   & 14.807   & 18\%           \\ \hline
2                       & 40                               & 1654, 991                                                  & 44.14    & 0.0062   & 0.462    & 1.072    & 49\%           \\ \hline
3                       & 68                               & 1654, 991, 1623                                            & 45.34    & 0.0054   & 0.363    & 0.955    & 45\%           \\ \hline
4                       & 192                              & 1654, 991, 1623, 1562                                      & 45.15    & 0.0055   & 0.317    & 0.992    & 75\%           \\ \hline
5                       & 378                              & 1654, 991, 1623, 1562, 1153                                & 46.33    & 0.0048   & 0.267    & 0.868    & 94\%           \\ \hline
6                       & 771                              & 1654, 991, 1623, 1562, 1153, 1083                          & 49.60    & 0.0033   & 0.222    & 0.590    & 94\%           \\ \hline
\end{tabular}}
\label{performance_metrics}
\end{table}

\begin{figure}
    \centering
    \includegraphics[width=3.5in]{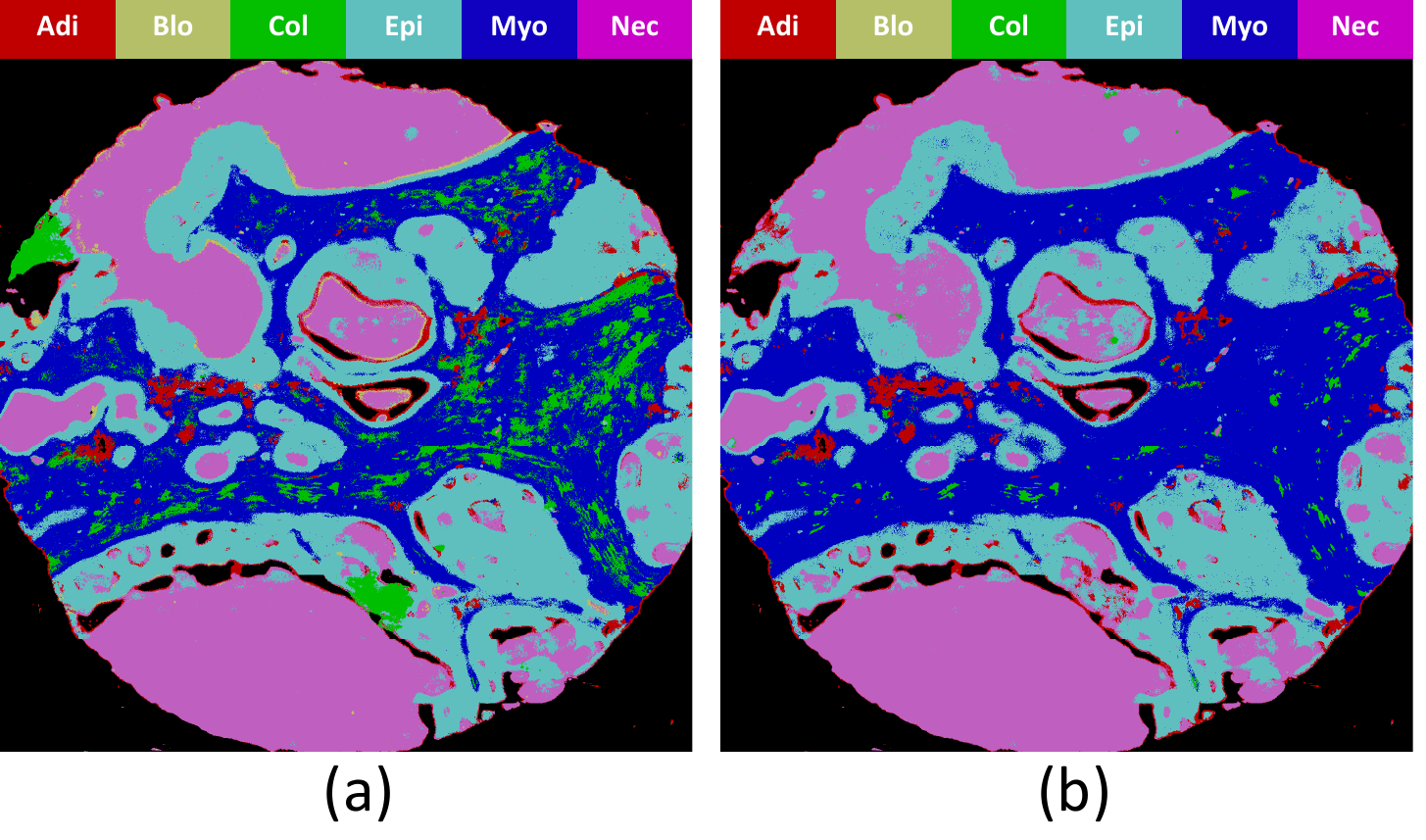}
    \caption{RBF SVM classification on the original FTIR breast core (a) and the reconstructed core using the proposed method (b).}
    \label{fig:svm_classification}
\end{figure}

\begin{figure}[tb]
    \centering
    \includegraphics[width=3.5in]{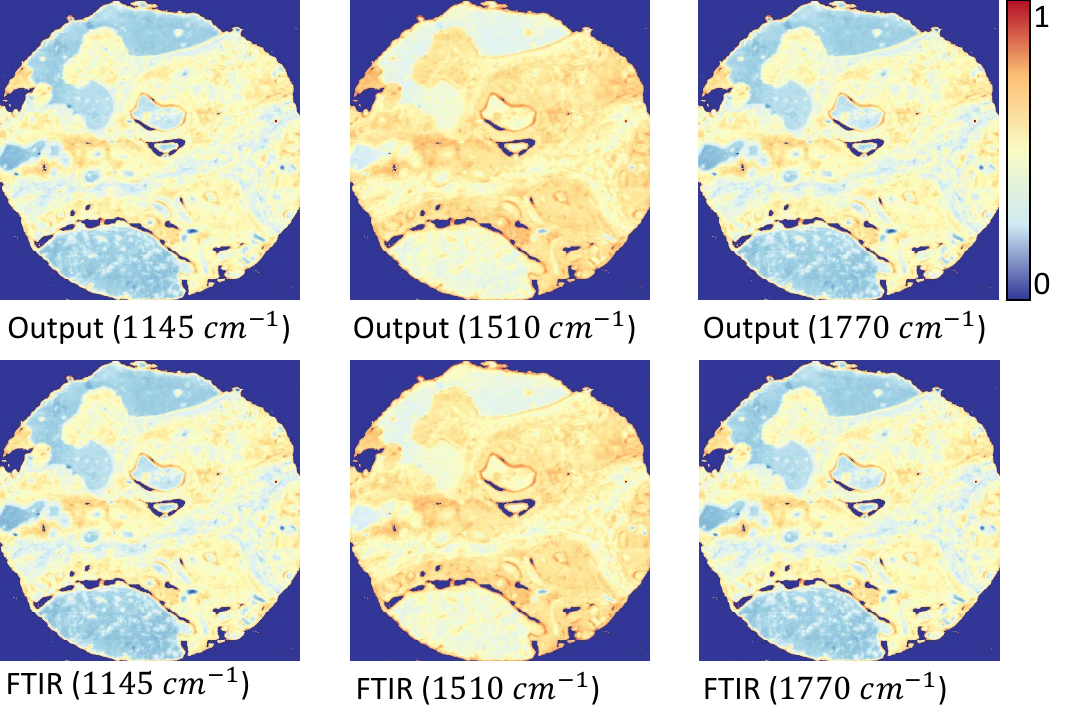}
    \caption{Infrared image of breast tissue biopsy in different wavenumbers. Top row: the reconstruction results using 5 band images and 378 point spectra. Bottom row: the original FTIR data (ground truth).}
    \label{fig:ftir_rec}
\end{figure}

The reconstructed data is evaluated using multiple, previously published metrics (Table \ref{performance_metrics}). Reconstruction after two iterations generates high-quality output based on  metrics such as PSNR and RMSE. However, the low classification accuracy indicates that the molecular content is not accurately reconstructed. After five iterations, using $5$ band images and 378 point spectra, the classification accuracy saturates  at $94\%$ and is comparable to the accuracy for raw data (Fig.\ \ref{fig:svm_classification}). The reconstructed and raw data in Fig.\ \ref{fig:ftir_rec} at multiple wavenumbers show good correspondence. Moreover, Table \ref{performance_metrics} shows that the reconstructed data at all  117 bands and 1,904,400 pixels is nearly identical to the raw data using multiple metrics. 

\section{Conclusion}
We propose an approach for reconstructing hyperspectral images using methods compatible with new high-resolution imaging instruments. Prior image generation methods \cite{li2017hyperspectral} require large datasets, and acquiring them will be prohibitive. Our proposed technique does not require the same quantity of training data as image generation methods and is therefore preferred. The reconstructions require less than 5\% of the original data by leveraging sparsity in both the spectral and spatial dimensions. The proposed algorithm selects the most informative individual spectra and band images. Sparse image reconstruction is done through weighted LASSO tensor completion. In addition to validating our approach using conventional metrics, we also propose using tissue classification as an alternative criterion. This approach more heavily weights the ability to distinguish critical molecular constituents, making the final reconstruction more useful for biochemical differentiation.

\bibliographystyle{styles/IEEEbib}
\bibliography{mirage-recon}

\end{document}